\documentclass[aps,prl,twocolumn,superscriptaddress]{revtex4}
\usepackage{graphicx}
\usepackage{amsmath}
\usepackage{amsfonts}
\usepackage[usenames]{color}
\usepackage{bm}
\newcommand{\bsy}[1]{\ensuremath{\boldsymbol{#1}}}

\newcommand{\bq}{{\bf q}}

\newcommand{\beq}{\begin{equation}}
\newcommand{\eeq}{  \end{equation}}
\newcommand{\bea}{\begin{eqnarray}}
\newcommand{\eea}{  \end{eqnarray}}
\newcommand{\bit}{\begin{itemize}}
\newcommand{\eit}{  \end{itemize}}

\begin{document}


\title{Observation of a Non-local Optical Vortex}

\author{R. M. Gomes}
\affiliation{Instituto de F\'\i sica, Universidade Federal de Goi\'as, GO 74.001-970, Brazil}
 \affiliation{Instituto de F\'{\i}sica, Universidade Federal do Rio
              de Janeiro, Caixa Postal 68528, Rio de Janeiro, RJ 21941-972,
              Brazil}
\author{A. Salles}
\affiliation{Instituto de F\'{\i}sica,
Universidade Federal do Rio de Janeiro, Caixa Postal 68528, Rio de
Janeiro, RJ 21941-972, Brazil}
\author{F. Toscano}
 \affiliation{Instituto de F\'{\i}sica, Universidade Federal do Rio
              de Janeiro, Caixa Postal 68528, Rio de Janeiro, RJ 21941-972,
              Brazil}
              \affiliation{ Funda\c{c}\~ao Centro de Ci\^encias e
              Educa\c{c}\~ao Superior a Dist\^ancia do Estado
              do Rio de Janeiro,
              20943-001 Rio de Janeiro, Brazil}

\author{P. H. Souto Ribeiro}
 \affiliation{Instituto de F\'{\i}sica,
Universidade Federal do Rio de Janeiro, Caixa Postal 68528, Rio de
Janeiro, RJ 21941-972, Brazil}
\author{S. P. Walborn}
\email[]{swalborn@if.ufrj.br}
\affiliation{Instituto de F\'{\i}sica, Universidade Federal do Rio
de Janeiro, Caixa Postal 68528, Rio de Janeiro, RJ 21941-972,
Brazil}

\date{\today}

\begin{abstract}
We report the observation of an optical vortex in the correlations of photons produced from spontaneous parametric down-conversion.  The singularity appears in a non-local coordinate plane consisting of one degree of freedom of each photon.    
 
\end{abstract}

\pacs{42.50.Xa,42.50.Dv,03.65.Ud}

\maketitle

The spatial correlations of photons have received much attention recently due to the interest in quantum information and fundamental aspects of quantum mechanics \cite{fonseca99b,howell04,dangelo04,walther04,neves05,nogueira04a}.  An important issue in this field is the investigation of  spatial entanglement through correlations of the orbital angular momentum (OAM) of photons \cite{mair01,franke-arnold02,caetano02,walborn04a,oemrawsingh05,molina-terriza07}.  In optics, OAM arises from a phase singularity in the transverse profile of a beam of light around which the phase fronts twist, and for this reason these light fields are also known as optical vortices or screw dislocations \cite{allen03}.  The motivation for the study of optical vortices ranges from fundamental interest to practical applications.  For example, OAM of optical vortex beams has been used to rotate micro-particles \cite{patterson01,curtis03}, and to produce coherent atomic vortices in Bose-Einstein condensates \cite{andersen06} and in cold atoms\cite{tabosa03}.    The OAM of single photons has been measured \cite{leach02} and has been proposed as a carrier of high-dimensional quantum information in quantum cryptography \cite{gibson04,spedalieri04,souza08} and quantum communication \cite{molina-terriza04,langford04,barreiro08}.  Furthermore, coherent transfer of OAM between light and matter opens the possibility of storage and retrieval of  high-dimensional quantum information \cite{barreiro03}.  
\par
Here we report the experimental observation of a non-local optical vortex associated with a phase singularity which manifests in the spatial correlations of two entangled photons.  The vortex appears in a two-dimensional plane of the two-photon phase space that is composed of one spatial degree of freedom of each photon.  The non-local character of this phase singularity makes it distinct from the usual OAM entanglement of pairs of photons, in which the pump laser or down-converted photons are prepared or projected onto a light mode possessing OAM. 
\par
A common example of beams with optical vortices are the Laguerre-Gaussian (LG) modes, which are a set of field modes with well-defined angular momentum.   The LG modes are parametrized by the radial and azimuthal quantum numbers $p$ and $m$, respectively.  A photon described by an LG mode carries an OAM equal to $m\hbar$.  The LG beams are usually created with holographic masks \cite{heckenberg92} or mode converters \cite{beijersbergen93}, the latter of which exploit the fact that a LG mode can be described as a superposition of Hermite-Gaussian (HG) modes.  For example,  a first-order ($p=0, m=\pm1$) LG mode can be written as a linear combination of HG modes \cite{beijersbergen93}
\begin{equation}
\mathcal{U}_{\mathrm{LG}}^{\pm}(x,y)=A\left [ u_{1}(x)u_{0}(y)\pm i u_{0}(x)u_{1}(y) \right],
\label{eq:LGsup}
\end{equation}
where the HG modes $\mathcal{U}^{\mathrm{HG}}_{kl}(x,y)=u_k(x)u_l(y)$ have been decomposed into the one-dimensional HG functions $u_{n}(x)$, given by the field amplitude 
\begin{align}
u_{n}(x)=& C_{n} H_{n}\left(\frac{x}{w(z)}\right)\exp\left(-\frac{x^2}{2 w(z)^{2}}
\right)\nonumber \\ & \exp\left\{-i\left[\frac{k x^2}{2R(z)}
-\left(n+\frac{1}{2}\right)\varphi(z)\right]\right\}.  
\label{eq:hg}
\end{align}
Here $C_{n}$ is a normalization coefficient,  $H_{n}(x)$ is the $n^{\mathrm{th}}$-order Hermite polynomial and the parameters $R(z)$, $w(z)$ and $\varphi(z)$ are the radius of curvature, beam waist, and Gouy phase, respectively.  We note that $u_{0}(x)$ is the usual one-dimensional Gaussian function.  
  \begin{figure}
\includegraphics[width=7cm]{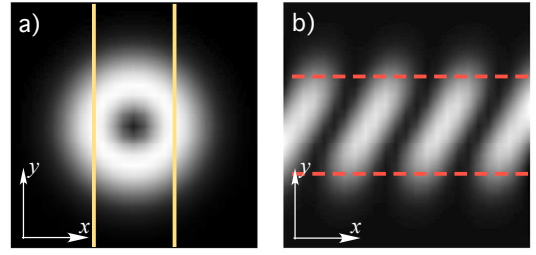}
 \caption{\label{fig:1}(Color online) a) Intensity profile of a first-order Laguerre-Gaussian mode.  b) Young's interference pattern from an infinitesimal double slit, represented by the vertical lines in a).  The fringe patterns is shifted, due to the phase structure of the Laguerre-Gaussian beam.  Along the horizontal dashed lines, the relative shift is half a period.}
 \end{figure}
\par

Let us now consider transverse spatial degrees of freedom of an entangled pair of photons. For simplicity, we use dimensionless position $\bsy{\rho}=(\rho_x, \rho_y)$ and wave vector $\bq=(q_x,q_y)$ variables.   In position space, the two photon wave function produced by spontaneous parametric down-conversion (SPDC) can be written as \cite{monken98a,walborn04a}
\begin{equation}
\psi\left(\bsy{\rho}_1,\bsy{\rho}_2\right) = \mathcal{E}(\bsy{\rho}_1+\bsy{\rho}_2) \Gamma(\bsy{\rho}_1-\bsy{\rho}_2), 
\end{equation}
where $\mathcal{E}(\bsy{\rho})$ is the field amplitude of the pump beam and $\Gamma(\bsy{\rho})$ is the phase matching function. Let us approximate the phase matching function by a Gaussian of width $\delta$: $\Gamma(\bsy{\rho})=\mathcal{U}_{00}(\bsy{\rho},\delta)=u_{0}(\rho_x, \delta)u_{0}(\rho_y, \delta)$, and consider that the pump laser is given by a first-order HG mode with width $\sigma$: $\mathcal{E}(\bsy{\rho})=\mathcal{U}_{10}(\bsy{\rho},\sigma)=u_{1}(\rho_x,\sigma)u_{0}(\rho_y,\sigma)$, where notation has been modified to include the widths explicitly.  Then, the two photon wave function is given by 
\begin{equation}
\psi\left(\bsy{\rho}_1,\bsy{\rho}_2\right) = \mathcal{U}_{10}(\bsy{\rho}_1+\bsy{\rho}_2,\sigma)  \mathcal{U}_{00}(\bsy{\rho}_1-\bsy{\rho}_2,\delta). 
\label{eq:state2} 
\end{equation}
Choosing $\sigma=\delta$, and using the fact that the HG functions are eigenstates of the Fourier transform, the two-photon state can also be described in a mixed position-momentum representation by the wave function 
\begin{align}
\psi\left(\bsy{\rho}_1,\bsy{q}_2\right) = &C  \left [ u_{1}(\rho_{x1})u_{0}(q_{x2}) + iu_{0}(\rho_{x1})u_{1}(q_{x2}) \right ] \nonumber \\
& \times u_{0}(\rho_{y1})u_0(q_{y2}), 
\label{eq:state2} 
\end{align}
where $C$ is a normalization constant.  
The $x$-component of the wave function is equivalent to the LG mode $\mathcal{U}_{\mathrm{LG}}^+$ given in Eq. \eqref{eq:LGsup}, where $\rho_{x1}$ and $q_{x2}$ play the role of the $x$ and $y$ variables.  Thus, the two-photon field in the $\rho_{x1} \times q_{x2}$ plane presents an optical vortex.    
The probability distribution $P(\rho_{x1},q_{x2})$, obtained by integrating $|\psi\left(\bsy{\rho}_1,\bsy{q}_2\right)|^2$ over $\rho_{y1}$ and $q_{y2}$, is 
\begin{equation}
\label{eq:state1a}
P(\rho_{x1},q_{x2})=|C|^2\left |u_{0}(\rho_{x1})u_{1}(q_{x2}) + iu_{1}(\rho_{x1})u_{0}(q_{x2}) \right|^2,  
\end{equation}
which corresponds to the modulus squared of the LG mode given in Eq. \eqref{eq:LGsup}, defined in non-local coordinates $\rho_{x1}$ and $q_{x2}$.

  \begin{figure}
\includegraphics[width=7cm]{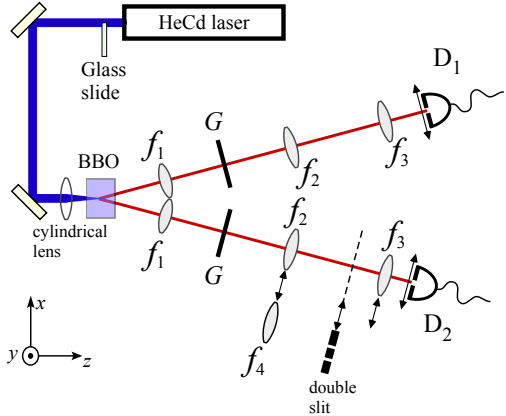}
 \caption{\label{fig:setup} (Color online) Experimental setup.  A non-local optical vortex is produced in the correlations of entangled photons produced by SPDC in a BBO crystal.  The two-photon state is engineered using the cylindrical lens to focus the pump beam and the Gaussian transmission masks (G) to spatially filter the down-converted photons.}
 \end{figure}
\par
Using photons produced from spontaneous parametric down conversion (SPDC), we generated a two-photon state accurately described by Eq. \eqref{eq:state2} and observed a non-local optical vortex.  The experiment is shown in FIG. \ref{fig:setup}.  A 441.6 nm c.w. HeCd laser is used to pump a 1cm long BBO crystal cut for type II phase matching.  The degenerate entangled photon pairs were  sent to single photon detectors outfitted with interference filters (40nm FWHM) centered at 884 nm.  Generation of the state \eqref{eq:state2} consists of three steps.  In the first step, a $125\mu$m thick microscope slide is inserted into half of the laser beam.  The angle of the slide is adjusted so that there is a $\pi$ phase shift between the two halves of the Gaussian beam.  After spatial filtering, the laser profile is approximately described by a HG mode $\mathcal{U}_{10}^{\mathrm{HG}}$.  In the second step, the pump laser is focused at the center of the BBO crystal using a cylindrical lens with focal length $33$mm.  This guarantees that the widths $\sigma$ and $\delta$ of the pump beam profile and phase matching function $\Gamma$ are approximately the same.  In the third step, the wave vector distributions of the down-converted photons at the center of the crystal are mapped onto intermediate planes through optical Fourier transforms ($f_1=100$mm).  The field profiles at the intermediate planes are spatially filtered using transmission masks $G$ with a Gaussian density profile, which guarantees the Gaussian shape of the function $\Gamma$.  The two-photon state at the intermediate planes just after the masks $G$ is given by Eq. \eqref{eq:state2}, where $\bsy{\rho}_1$ and $\bsy{\rho}_2$ are the transverse position coordinates.  
  \begin{figure}
\includegraphics[width=7cm]{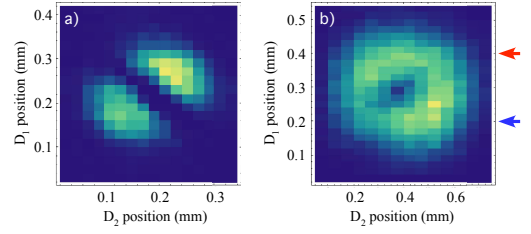}
 \caption{\label{fig:results} (Color online) a) Coincidence count distributions corresponding to the probability distribution $P(\rho_{x1},\rho_{x2})$ are described by a Hermite-Gaussian mode.  b) Coincidence count distributions corresponding to the probability distribution $P(\rho_{x1}, q_{x2})$ are described by a Laguerre-Gaussian mode.  The zero intensity region at the center is due to the undefined azimuthal phase at the origin.  The arrows show the positions of detector 1 used to obtain FIG. \ref{fig:results2}.}
 \end{figure}
\par
To verify the presence of the HG mode in the entangled two-photon wave function, the intermediate planes are mapped onto the detection planes using an imaging system composed of two confocal lenses with focal lengths $f_2=150$mm and $f_3=50$mm.  The $\rho_{x1} \times \rho_{x2}$ cut of the phase space distribution is measured by scanning the two detectors in the vertical direction, resulting in a two-dimensional (2D) array of coincidence counts that is proportional to the detection probability $P(\rho_{x1},\rho_{x2})$, as shown in FIG. \ref{fig:results} a). In this figure one can see that the coincidence count distribution along the diagonal, which displays correlations coming from the function $\mathcal{E}(\rho_{x1} + \rho_{x2})$, is indeed given by first-order Hermite-Gaussian mode.  The anti-diagonal direction corresponds to correlations coming from the function $\Gamma(\rho_{x1} - \rho_{x2})$, and is approximately described by a Gaussian function.   Next,  the Fourier transform of the intermediate plane $G$ of photon 2 is performed using a lens with focal length $f_4=250$mm.  Since the detection plane lies in the focal plane of this lens, the field distribution in the detection plane corresponds to the Fourier transform of the plane $G$ up to a quadratic phase term, which does not contribute to the intensity distribution.  {FIG. \ref{fig:results} b) shows the 2D array of coincidence counts corresponding to the probability distribution $P(\rho_{x1}, q_{x2})$, which displays the same doughnut shape that is characteristic of the intensity profile of the LG field mode, as shown in FIG. \ref{fig:1} a).  The coincidence distribution is not a perfect replication of FIG. \ref{fig:1} a), due to noise which is most likely due to slight misalignment of the lenses, the Poissonian photon count statistics, and the finite width of the detector apertures ($20\mu$m for $\rho$ measurements and $50\mu$m for $q$ measurements).  Despite this noise, the doughnut shape, in particular the zero intensity region at the center, is clearly visible.}               
\par
  \begin{figure}
\includegraphics[width=7cm]{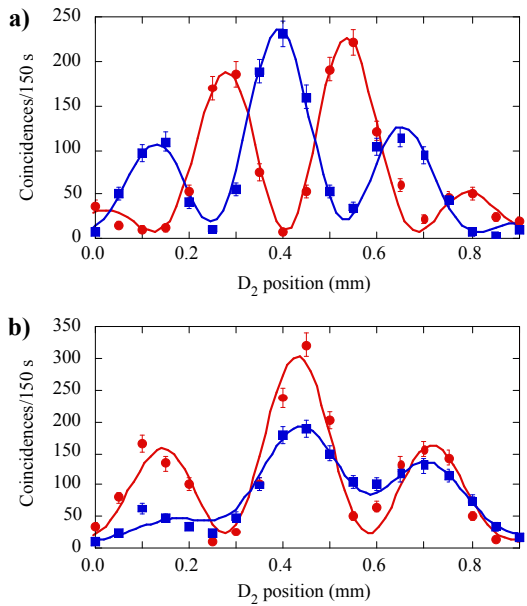}
 \caption{\label{fig:results2} (Color online) Interference fringes when the pump beam is a) a Hermite-Gaussian beam and b) a Gaussian beam.  In both figures the sets of fringes are obtained by fixing detector 1 at positions $\rho_{x1}=0.2$ (red circles) and $\rho_{x1}=0.4$ (blue squares) and scanning detector 2.  Solid lines correspond to curve fits of Eq. \eqref{eq:inter}.}
 \end{figure}
Even though the coincidence profile in FIG. \ref{fig:results} b) resembles the intensity profile of a LG mode, in order to verify the existence of a phase singularity it is necessary to characterize the phase dependence.  A simple method to identify the  azimuthal phase dependence of an optical vortex is through interference in a typical Young's double slit experiment \cite{sztul06},  as illustrated in Fig. \ref{fig:1} b).  The intensity of the Fraunhoffer diffraction pattern of a Laguerre-Gaussian beam incident on a double slit, oriented to produce fringes in the $x$ direction, is given by
\begin{equation}
I(x,y) = R(x,y)\left[ 1 + \cos\left (\frac{\alpha(x) + \Delta\phi(y)}{2} \right) \right],
\label{eq:inter}
\end{equation}   
where $R(x,y)$ is a Gaussian function,  $\alpha(x)$ is the usual phase difference due to propagation from the each slit to the detection plane and $ \Delta\phi(y)=\phi_1(y)-\phi_2(y)$ is the phase difference of the input LG field at each slit.  We emphasize that the interference fringes now depend on both $x$ and $y$ directions.  The azimuthal phase structure results in an oblique set of fringes as illustrated in FIG. \ref{fig:1} b), in contrast to the usual vertical linear fringes which appear in Young's interference of a plane wave or Gaussian mode.  Therefore, the fringe pattern obtained by scanning in the $x$ direction for different $y$ positions will be phase-shifted from one another.
Choosing the $y$ positions represented by the dashed lines in FIG. \ref{fig:1} b), the shift can be close to half a period.       
\par
To characterize the azimuthal phase structure,  photon 1 was measured using the imaging system, while a double slit (200$\mu$m slits, 100 $\mu$m separation) was placed in the focal plane of lens $f_2$ of photon 2.  The field distribution in this plane corresponds to the wave vector distribution of the field at the plane of the mask G.  Lens $f_3$ was then used to map the Fourier transform of the field distribution along the double slit onto the detection plane of detector 2, corresponding to the Fraunhofer diffraction regime.   Detector $D_1$ was placed in two positions ($0.2$mm and $0.4$mm) indicated by the arrows in FIG. \ref{fig:results} b), while detector 2 was scanned linearly in the transverse plane, corresponding to the horizontal axis in FIG. \ref{fig:results} b).  The result was the set of interference fringes shown in FIG. \ref{fig:results2} a).  A relative shift of approximately half a period between the two sets of fringes is clearly observed.  For comparison, we removed the glass microscope slide from the pump laser, obtaining a Gaussian profile beam.  The corresponding interference fringes are shown in FIG. \ref{fig:results2} b).  In this case no shift of fringes is observed, demonstrating that the features in FIG. \ref{fig:results2} a) are indeed a consequence of the azimuthal phase dependence of the spatial correlations.  The asymmetry of the fringes in FIG. \ref{fig:results2} b) is due to a small distortion in the shape of the Gaussian beam in the non-local plane. 
\par
 It is interesting that the non-local vortex appears in the coordinates of entangled photons, which suggests that there may be a relation between entanglement and isolated phase singularities.  We will now show that appearance of an isolated phase singularity implies that a bipartite pure state is entangled.  Consider the wave function $\phi(x,y)=|R(x,y)|\exp{[ih(x,y)]}$, where $x$ and $y$ are coordinates for two modes.  Let us choose the vortex such that $h(x,y)$ is undefined at the origin.  For $\phi$ to be separable requires $h(x,y)=h_1(x)+h_2(y)$, which implies that $h_1$  and/or $h_2$ is undefined at $x=0$ and/or $y=0$, respectively.  But this implies that $\phi(x,y)$ is singular at all points along the line $x=0$ and/or $y=0$, which contradicts the original assumption of an isolated singularity.  Thus, in the present case the observation of the non-local vortex implies in quantum entanglement.
\par           
 Let us briefly discuss our results in terms of propagation of the two-photon state, considering one spatial dimension of each photon.  At the source the spatial profile of coincidence counts is given by the diagonal HG mode shown in FIG. \ref{fig:results}a).  After photon 1 passes through an imaging system and photon 2 through a Fourier transform system, the coincidence profile is given by  the LG mode shown FIG. \ref{fig:results}b).  This propagation is analogous to the mode conversion of a classical HG beam to a LG beam, achieved through introduction of an astigmatic propagation region \cite{beijersbergen93}.  Here  the astigmatism occurs between the propagation of photons 1 and 2,  and corresponds to the photons propagation through different lens systems:  the imaging system of photon 1 and the Fourier system of photon 2.  In this sense, propagation of the two-photon state from the Gaussian masks to the detection plane can be considered as a non-local mode conversion. 
 \par
In conclusion, we have observed a phase singularity in a non-local coordinate plane of a pair of entangled photons.  The phase singularity appears as a non-local optical vortex in the spatial correlations of the entangled photons.  Though we have shown the existence of a first-order vortex, it should be possible to create higher-order vortices, or more complicated structures.  We also note that these non-local singularities should appear in any pair of entangled harmonic oscillators, such as modes of the electromagnetic field or the vibrational modes of trapped ions.         
One might envision using these types of vortices to encode quantum information non-locally in two photons or two atomic clouds, or to produce a two-photon angular momentum transfer effect in two-photon absorption.

\begin{acknowledgements}

Financial support was provided by Brazilian agencies CNPq,
CAPES, FAPERJ, and the Milenium Institute for Quantum
Information.  We thank D. P. Caetano for bringing Ref. \cite{sztul06} to our attention. 
\end{acknowledgements}




\end{document}